# Plasmon drag effect with sharp polarity switching

T. Ronurpraful, D. Keene, N. Noginova

*Norfolk State University, Norfolk VA*
Email: d.w.keene@spartans.nsu.edu

The generation of significant photocurrents observed in plasmonic metasurfaces is interesting from a fundamental point of view and promising for applications in plasmon-based electronics and plasmonic sensors with compact electrical detection. We show that photoinduced voltages in strongly modulated plasmonic surfaces demonstrate a highly asymmetric angular dependence with polarity switching around the plasmon resonance conditions. The effects are tentatively attributed to coupling between localized and propagating plasmons.

**Introduction**

Significant photoinduced electric effects observed in metal films and nanostructures [1-13] are shown to be closely associated with excitation and propagation of surface plasmons. The amplitude of the electric currents observed in the plasmonic surfaces at plasmon resonance conditions can exceed the predictions of a simple photon drag (light pressure [14, 15]) mechanism by orders of magnitude [2-4]. This plasmon drag may provide an opportunity to monitor plasmonic excitations electrically and is promising for applications in plasmonic circuits [16-18], optoelectronics with plasmonic elements [19], and biomedical sensors with compact electric detection [20]. On the other hand, generation of photocurrents in plasmonic surfaces is interesting from a fundamental point of view as it relates to the processes of energy transfer, linear momentum transfer, and angular momentum transfer between light and matter in plasmonic systems.

The origin of the plasmon-enhanced electric effects is still not fully clear. Typically, the polarity of the signal corresponds to electron drift in the direction of surface plasmon polariton (SPP) propagation. These observations are reasonably well described with the electromagnetic momentum loss approach [21-23] operating in terms of effective forces ("plasmonic pressure" and "striction") associated with strong optical fields and strong gradients of the optical fields at the resonance conditions. Plasmon drag is viewed as the result of momentum exchange between SPPs and electrons (plasmonic pressure mechanism) [22]. An additional assumption is required to correctly describe the magnitude of the signals; it was assumed that the momentum relaxation time of free electrons is not the Drude's constant but corresponds to the energy relaxation time [22]. However, this relatively simple approach does not describe the full range of the effects observed in various experiments in nanostructured surfaces. One such effect is the presence of electric signals of a smaller magnitude and the opposite polarity, corresponding to the electron drag against the SPP k-vector and against the k-vector of incident photons [2, 7].

Recent studies [24] of photoinduced electric currents in flat gold films with very low roughness (<1 nm) in a vacuum yield intriguing results, calling for a reassessment of the theoretical approaches to the plasmon drag phenomenon. Similar to previous findings [1-2], at ambient conditions significant photocurrents are observed at the SPP resonance with the polarity parallel to the SPP k-vector. However, when the air is pumped off, the signal peak at the SPP resonance strongly decreases in magnitude and changes its polarity to the opposite. This corresponds to a drift of electrons against the SPP k-vector in clear contradiction to the simple plasmon-to-electron momentum transfer mechanism.

In [7], it was shown that in plasmonic gratings with smooth modulation profile and relatively low modulation amplitude, photocurrents correspond to electron drift in the direction of SPP propagation, and can be reasonably well described with the "plasmonic pressure" mechanism [21-22]. In the current work we use plasmonic gratings with steep edges and a relatively high amplitude profile modulation, and observe quite unusual behavior of the photocurrents. We also explore a possible application of these effects. We believe that our results can add more information towards a better understanding of the origin of plasmon-related electric effects and their practical applications.

**Experimental**

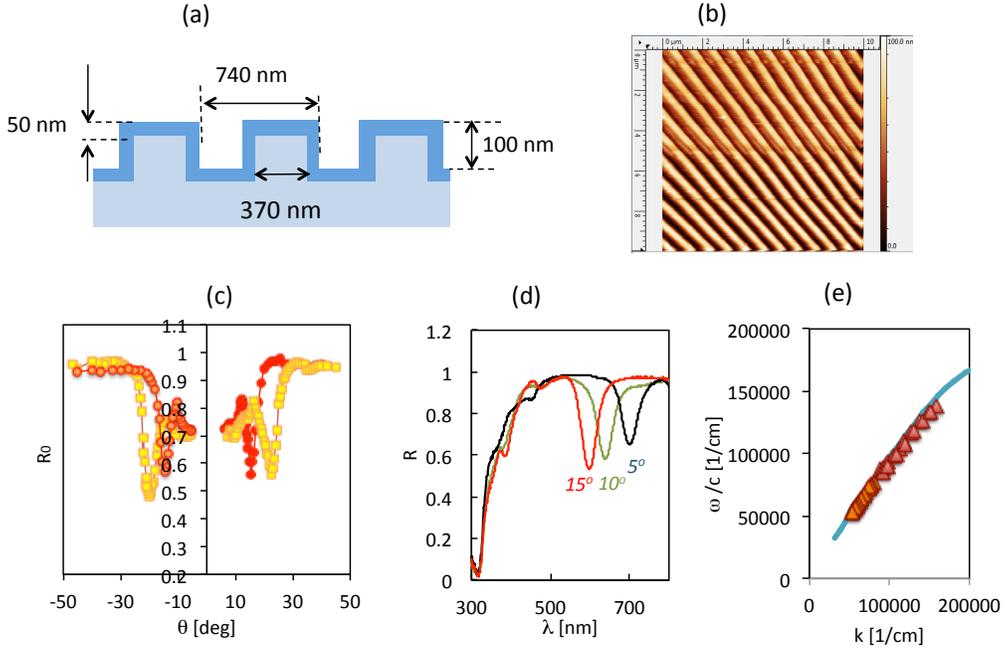

Fig. 1. (a) Schematics and (b) AFM image of Ag/DVD structures; (c) Reflectivity v*s* angle for $\lambda$ = 585 nm (yellow) and 632.8 nm (red); (d) Reflection spectra at p polarization and various orientations as indicated; (e) SPP dispersion relationship, Ag/DVD (points) and estimation for flat films, Eq. 2 (solid trace).

We have found that convenient systems for our experiments can be derived from commercial DVD disks. Polycarbonate substrates obtained from disassembled DVD-Rs (manufactured by Verbatim) have concentric periodic grooves with a periodicity of 740 nm. The details of the disk disassembling and cleaning process are described in [25]. Since the radius of these concentric circles is much higher than the other characteristic parameters in our experiments (the grating period, surface plasmon polariton (SPP) propagation length and size of the illumination spot), the presence of the groove curvature is not expected to greatly affect the results. A large area of the obtained surface allows one to cut the substrates for experimental samples with the desired geometry and orientation of the grooves. Silver, gold or other metals with a thickness of ~50 nm is deposited on the precut substrates with the thermal vapor deposition technique. We also use structures with the original metal layer (which is silver, according to Scanning Electron Microscopy integrated with Energy Dispersive X-Ray Spectroscopy (SEM-EDS) and X-Ray Diffraction (XRD) analysis and optical characterization). All the systems tested show similar angular dependences of the photoinduced electric effects (however, of a different magnitude). Here we concentrate on the results obtained in silver Ag/DVD systems. Also, we should note that a similar procedure can be applied for fabrication of metal gratings from BluRay-R (BR) discs. However, since the obtained BR systems have a smooth modulation profile with a relatively small modulation height (~ 25 nm), the plasmon drag experiments yield results very similar to those previously reported [7], and are not presented here for the sake of brevity.

The schematics and Atomic Force Microscopy (AFM) image of the Ag/DVD structure are shown in Fig. 1(a) and (b) respectively. This is basically a metal grating with deep grooves, which have nearly rectangular profile and a height of ~ 100 nm. In periodically modulated metal surfaces, SPPs are excited with direct illumination at the coupling condition [26],

$$k_{SPP} = mG + k_x, \qquad \textbf{(1)}$$

where $m \neq 0$ is an integer, $G = 2\pi/d$ is the grating vector, $d$ is the period of modulation, $k_x = k_0 \sin\theta$ is the projection of the optical k-vector ($k_0$) onto the metal and dielectric interface plane, $\theta$ is the angle of incidence, and $k_{spp}$ is the SPP wave-vector. In flat films [26],

$$k_{SPP} = \frac{\omega}{c}\sqrt{\frac{\varepsilon_m \varepsilon_d}{\varepsilon_m + \varepsilon_d}}, \qquad (2)$$

where $\omega$ is the light frequency, and $\varepsilon_m$ and $\varepsilon_d$ are correspondingly dielectric permittivities of metal and dielectric.

The excitation of SPPs in Ag/DVD structures is seen as dips in the angular dependence (Fig. 1 (c)) of the reflectivity under illumination with p-polarization. In the experiment shown in Fig.1 (d), the sample is placed in the center of the integration sphere, and the reflected intensity is measured *vs* wavelength at different sample orientations at the horizontal plane. The p-polarized spectra show well defined dips due to the absorption associated with SPPs. The dispersion relationship $\omega(k)$ derived from the spectral dependence of the dip positions and Eq. 1 (at $m = 1$) is shown in Fig 1 (e). The small discrepancy with the predictions of Eq. (2) (solid trace) can be ascribed to a relatively high profile modulation. Additional shallow dips observed at the spectral dependences of Fig. 1 (d) can be related to -2d order SPPs (m = -2) or localized surface plasmons (LSPs).

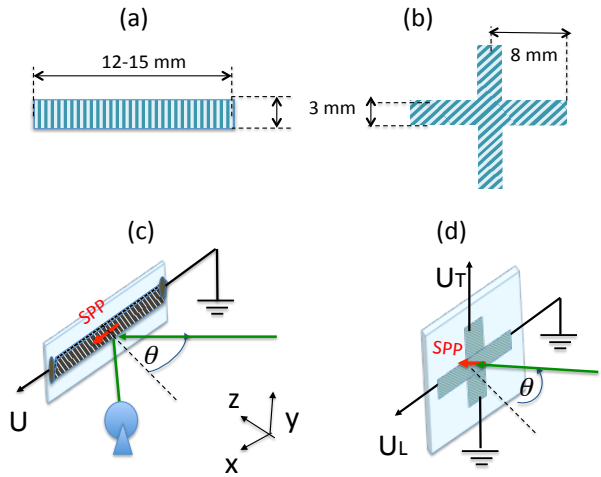

Fig. 2. (a) and (b) Schematics of the experimental samples, single strip (a) and plus-shaped (b). (c) and (d) experimental setups and geometry of the measurements for strip and plus shaped samples correspondingly.

Schematics of the samples and the experimental setup are shown in Fig. 2. Most samples are cut in the shape of a strip with a width of ~ 3 mm and length 12-15 mm, Fig. 2 (a). The grooves are oriented perpendicular to the long side. The electric contacts are attached on either end, allowing one to record the voltage generated along the strip (Fig. 2 (c)). Another geometry used is the shape of plus sign, with grooves oriented ~ 45 degree to both arms, Fig. 2 (b). Contacts are placed on each set of arms for recording voltages in longitudinal (in the direction of the incidence plane) or transverse (perpendicular to the incidence plane) directions, $U_L$ and $U_T$ respectively (Fig. 2 (d) The sample is placed on a goniometer stage and illuminated with an Optical Parametric Oscillator (OPO) at various wavelengths at p-polarization. The width of the illuminated area is slightly larger than the width of the strip. The duration of the pulse is ~ 5 ns, and the pulse energy is 0.3-0.5 mJ per pulse. The photoinduced voltage is recorded with a digital Tektronix oscilloscope with 50 Ω internal resistance as a function of the angle of incidence, $\theta$. In order to check the position of the SPP dip, the reflected intensity is collected with an integration sphere.

Typical results are shown in Fig. 3. As expected, the electric signals in the structures are strongly enhanced at the angle corresponding to the SPP excitation, $\theta_{spp}$. However, in contrast to the previous studies [1, 2, 7], the SPP-related enhancement has a highly asymmetric Fano like shape, Fig. 3 (a). At incidence angles lower than $\theta_{spp}$, the polarity of the voltage (negative in the graph) corresponds to the electron drag along the SPP k-vector. It abruptly switches to the opposite polarity at the angles above the resonance angle, and remains positive with quickly decreasing magnitude with a further increase in $\theta$. (Note that in all experiments presented here, we use p-polarized light. Illumination with s-polarization produces voltages as well but of much smaller magnitude and without any polarity switching. The results obtained at s-polarization will be presented elsewhere.)

Further experiments confirm the sharp polarity switching at various wavelengths of illumination, Fig. 3 (b, c) with different positions of the switching point. In all cases, the switching angle approximately corresponds to the SPP resonance angle, Fig. 3 (d). The angular width and relative magnitudes of parallel and antiparallel components vary among samples, particularly depending on the freshness of silver films. In freshly made samples (Fig. 3, a and b), the parallel component is significant while in deteriorated (one – two weeks old) samples (Fig. 3 (c)), the antiparallel component is prevailing.

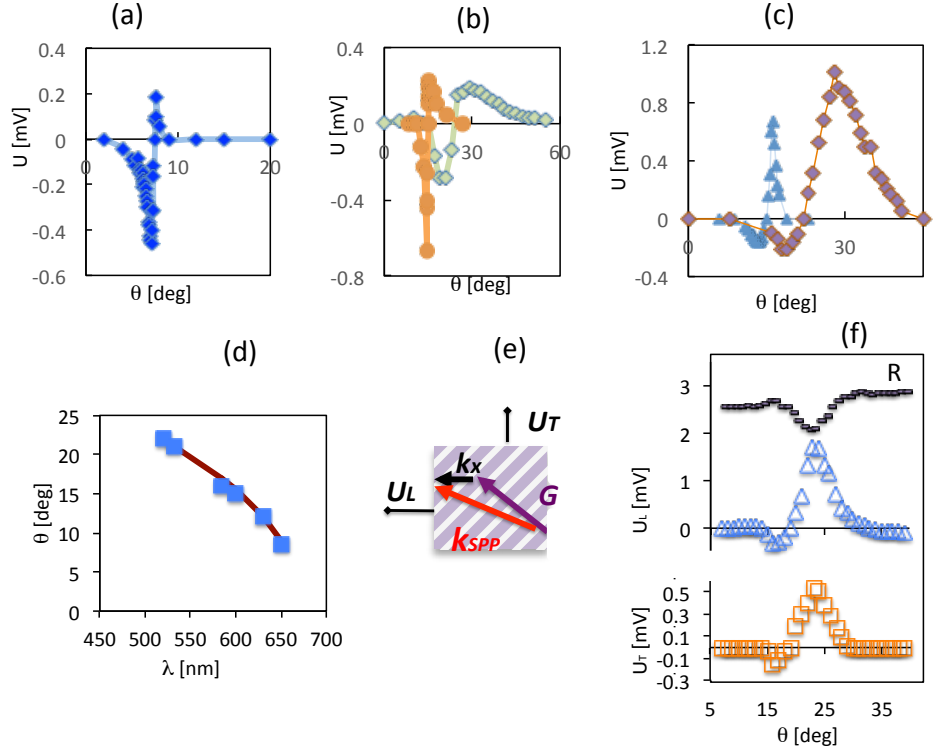

Fig. 3. (a-c) Switching polarity in Ag/DVD strip-shaped samples. Negative polarity corresponds to the electron drift in the direction of SPP propagation. (a,b) $U(\theta)$ in freshly made samples at $\lambda$ = 650 nm (a), $\lambda$ = 600 nm (orange circles) and 532 nm (green diamonds) (b). (c) $U(\theta)$ in the two-week old sample, $\lambda$ = 585 nm (blue triangles) and 520 nm (brown diamonds). Solid traces are guide for eyes. (d) Spectral dependence of the switching angle (points), and the spectral position of the SPP related dip (solid trace). (e) Schematics of the SPP excitation and (f) longitudinal (blue triangles) and transverse (orange squares) voltages in plus-shaped sample at $\lambda$=585 nm. The corresponding dip in reflected intensity, $R$, is shown in black.

This switching phenomenon is surprising and is not readily predicted with the plasmonic pressure model. A question arises as to whether the observed switching phenomenon is driven purely by SPP propagation or if it can be the result of an interplay between the propagating plasmon and the incident radiation. In order to test this, we use the plus-shaped sample, with grooves oriented at 45° with respect to the incidence plane, and simultaneously record the longitudinal voltage $U_L$ induced in the direction of $k_x$ and transverse voltage, $U_T$ induced in the perpendicular direction. The excitation of the SPP is described with the same Eq. 1 but in vector form. The SPP is excited under ~ 33° with respect to the direction of $k_x$, Fig 3 (e). In this case, if the observed switching is due the interplay of incident light and SPP, it will be observed only in $U_L$. As one can see in Fig. 3 (f), the switching is observed in both longitudinal and transverse voltages, with almost the same shapes of the angular dependence. The switching point is close to the SPP resonance angle, see the dip in the simultaneously recorded reflected intensity in Fig 3 (f). This result confirms the direct relationship between the switching and the SPP excitation.

In the second series of the experiments, we test the dependence of the effect on the dielectric environment. According to Eqs.1 and 2, a change in the dielectric constant of the environment results in a shift of the SPP resonance angle. This effect is commonly used in plasmonic-based sensing applications [27-31]. As one can see in Fig. 4 (a, b), adding a polystyrene layer (with the refractive index of 1.6 and thickness of ~ 800 nm) on the top of the Ag/DVD structure results in a shift of the switching point to higher angles while the switching phenomenon occurs in both air and polymer environments. Assuming standard values of the permittivity for silver [32] and the permittivity of dielectric as 1 (air) or 2.6 (polymer), the estimations yield $\theta_{SPP} = 9°$ and $59°$ in air and polymer correspondingly, which are close to the observed switching positions.

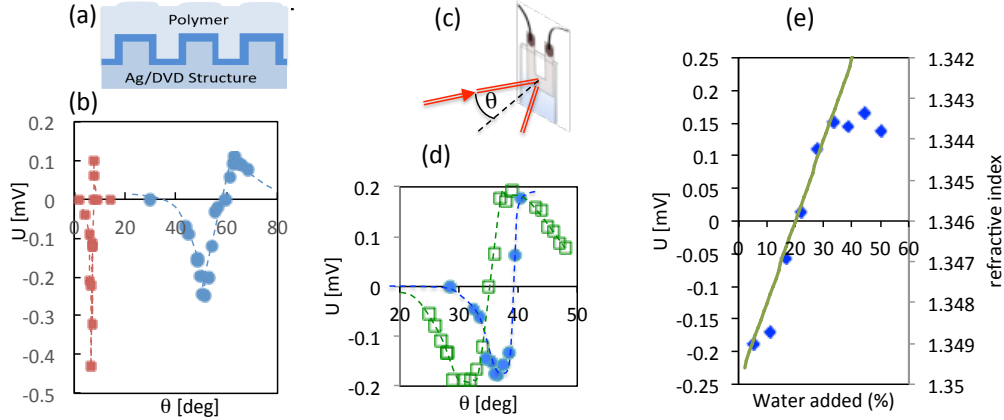

Fig. 4. (a) Schematics of the Ag/DVD with polymer layer, (b) $U(\theta)$ without (red) and with polymer layer (blue). Dashed lines are guide for eyes; (c) Schematics of the experiment with ethanol-water solution; (d) $U(\theta)$ in 70 % ethanol (blue) and water (green), $\lambda = 650$ nm; (e) Voltage recorded at $37.5°$ as a function of water addition (points) and corresponding change in the refractive index (solid trace). $\lambda = 630$ nm.

Note that Fano resonances in plasmonic systems present a special interest for plasmonic-based sensing applications [31] as they are associated with sharp features in optical reflectivity and may significantly enance the capability to resolve small changes in the dielectric constant of the environment and figure of merit of the sensors. However, optical monitoring of common plasmonic sensors requires a bulk optical setup, but electric reading can provide compact plasmonic devices for such applications. In the experiment shown in Fig. 4 (c-e), we study the feasibility to use the polarity switching in Ag/DVD structures for electric detection of small changes in the refractive index. The sample prepared in a U-shape (Fig. 4 (c)) is placed in a water-ethanol environment, where the refractive index can be gradually changed by changing the relative concentrations of water and ethanol. Since the dielectric constants of ethanol and water are different (the refractive indexes are 1.35 [33] and 1.33 [34] respectively) the switching positions of the photocurrent polarity are different as well (Fig. 4 (d)).

In the experiment shown in Fig 4 (e), the sample in 70 % ethanol is oriented at a fixed angle of $37.5°$ corresponding to the negative polarity of the electric signal. Small volumes of water are added changing the relative concentration of the solution by 5.5% by volume with each step. After a few steps, the PLDE signal switches its polarity to positive, since the switching point shifts toward lower angles with a higher concentration of water. Assuming that the dielectric constant of the water/ethanol solution is the sum of the dielectric constants of water and ethanol with the corresponding weight, we estimated the modification in the refractive index of the environment during the experiment (solid trace in Fig. 4 (e)). As one can see, $\Delta n = 0.001$ is resolved with the change in the photocurrent polarity.

**Discussion**

Let us summarize the experimental findings: The photoinduced electric currents in our structures are strongly enhanced at SPP resonance. However, in contrast to previous studies in flat films and gratings with smooth profiles and low modulation height [1, 2, 4, 7], the angular dependence of the enhanced photocurrents is not a single peak corresponding to the drift of electrons parallel to SPP. It has a highly asymmetric shape with sharp switching in the photocurrent polarity from the drift of electrons together with the SPP at low angles to the drift of electrons against the SPP at high angles. One can consider the possibility that it is simply an interplay between 1st order and –2nd order SPPs. However, the experiments at various wavelengths and in different dielectric environments indicate that the switching point is directly related to the 1st order SPP, which has a pronounced optical signature in the reflectivity as a single dip. In this case, the plasmonic pressure model predicts electron drift only in one direction corresponding the direction of SPP propagation.

In order to explain our observations, at least tentatively, let us take a different approach to plasmon-related electric effects, and assume that in strongly nanostructured plasmonic surfaces (such as rough films, nanomesh structures, arrays of nanopillars or gratings with high profile modulation magnitudes) nanosize features are the main sources of the photoinduced electric signals rather than propagating SPPs. Such a hypothesis is used in [3], in order to explain the spectral dependence of the plasmon drag effect in nanostructured systems (where the signal has a maximum at the wavelength of the localized plasmon (LSP) resonances of nanosize features). There are other findings, which are in line with this hypothesis: Photoinduced electric voltages in nanostructured systems are found to be strongly dependent on the nanoscale geometry of surface features [3]. Nanostructures significantly enhance the magnitude of photoinduced voltages [3, 7, 11, 13], and even a slight asymmetry of the nanosize features [3] defines the photocurrent polarity independent of incidence angle. In [35], the Kelvin probe technique was used to record photoinduced electric potentials in metal nanostructures. Under CW illumination, negative or positive potentials are generated, depending on whether the wavelength of illumination is below or above the LSP resonance.

Note that the shape of the angular dependence of the photovoltage in our structures strongly ressembles the Fano resonance shape [36-38], and may result from the coupling of two plasmon resonances, one with a broad and one with a narrow linewidth. In our structures, the narrow mode is the 1st-order SPP, and the broad resonances can be LSPs excited at the sharp corners of the structure or at other defects. According to our current consideration, the SPPs are not the main source of the photoinduced voltage but rather a driving force, which controls voltages generated in the individual sources. In a similarity with [35] where the switching in relative potentials is observed at the LSP conditions, our photocurrent polarity switching is observed at the resonance of the SPP-LSP coupled system. An increase in photocurrent magnitudes in slightly deteriorated structures (Fig. 3 (c)) may be related to increase in the sample roughness and higher number of participating "voltage sources".

Possible mechanisms of the voltage generation in nanofeatures may include thermoelectric effects [39], nonlinearity and asymmetry of electron motion [3, 40] or charging due to high field gradients, which can be described theoretically [21] in terms of the effective striction force. The striction force is conservative [22]; it is expected to affect only the electron density distribution and not result in the total photocurrent across the sample even in the case of highly asymmetric intensity distribution around a nanofeature. However, if the size of the nanofeature is shorter than the electron propagation length, the ballistic regime of electron motion comes into play, potentially contributing to the resulting current.

In conclusion, photocurrents in plasmonic gratings with a nearly rectangular modulation profile sharply switch their polarity around the SPP resonance angle. The switching is shown to be sensitive to the dielectric environment and can find applications in plasmonic based sensing. The results are discussed assuming of the major role of the nanoscale features in voltage generation and coupling between SPPs and LSPs.

Acknowledgements:
Authors would like to acknowledge financial support from National Science Foundation (NSF) (1646789, 1830886); Air Force of Scientific Research (AFOSR) (FA9550-18-1-0417) and Department of Defense (DoD) (W911NF1810472).

### References
[1] Vengurlekar A and Ishiara T 2005 Surface plasmon enhanced photon drag in metals *Appl. Phys. Lett.* **87** p 091118
[2] Noginova N, Yakim V, Soimo J, Gu L, and Noginov M A 2011 Light-to-current and current-to-light coupling in plasmonic systems *Phys. Rev. B* **84** p 035447
[3] Noginova N, Rono V, Besares F J and Caldwell J D 2013 Plasmon drag effect in metal nanostructures *New J. Phys.* **15** p 113061
[4] Kurosawa H and Ishihara T 2012 Surface plasmon drag effect in a dielectrically modulated metallic thin film *Opt. Express* **20** p 1561

[5] Kurosawa H, Ishihara T, Ikeda N, Tsuya D, Ochiai M, and Sugimoto Y 2012 Optical rectification effect due to surface plasmon polaritons at normal incidence in a nondiffraction regime *Opt. Lett.* **37** p 2793
[6] Bai Q 2015 Manipulating photoinduced voltage in metasurface with circularly polarized light *Opt. Express* **23** p 5348
[7] Noginova N, LePain M, Rono V, Mashhadi S, Hussain R, and Durach M 2016 Plasmonic pressure in profile-modulated and rough surfaces *New J. of Physics* **18** 093036
[8] Ishihara T, Hatano T, Kurosawa H, Kurami Y, and Nishimura N, 2012 Transverse voltage induced by circularly polarized obliquely incident light in plasmonic crystals *Proc. SPIE Spintronics V*
[9] Ni X, Xiao J, Yang S, Wang Y, and Zhang X 2015 Photon induced collective electron motion on a metasurface *Conference on Laser and Electro-Optics*
[10] Kang L, Lan S, Cui Y, Rodrigues S P, Liu Y, Werner D H, and Cai W 2015 An active metamaterial plarform for chiral responsive optoelectronics **27** p 4377
[11] Akbari M, Onoda M, and Ishihara T 2015 Photo-induced voltage in nano-porous gold thin film *Opt. Express* **23** p 823
[12] Mikheev G M, Saushin A S, and Vanyuko V V 2015 Helicity-dependent photocurrent in the resistive Ag/Pd films excited by IR laser radiation *Quantum Electron* **45** p 635
[13] Noginova N, Rono V, Jackson A, Durach M, 2015 Controlling plasmon drag with illumination and surface geometry *Conference on Laser and Electro-Optics*
[14] Gibson A F, Kimmitt M F, and Walker A C 1970 Photon drag in germanium *Appl. Phys. Lett.* **17** p 75
[15] Loudon R, Barnett S M, and Baxter C 2005 Radiation pressure and momentum transfer in dielectrics: the photon drag effect *Phys. Rev. A* **71** p 063802
[16] Engheta N 2007 Circuits with light at nanoscales *Science* **317** p 1698
[17] Adato R, Yanik A A, and Altug H 2011 On chip plasmonic monopole nano-antennas and circuits *Nano Lett.* **11** p 5219
[18] Liu N, Wen F, Zhao Y, Wang Y, Nordlander P, Halas N, and Alu A 2013 Individual nanoantennas loaded with three-dimensional optical nanocircuits *Nano Lett.* **13** p 142
[19] Kang L, Lan S, Cui Y, Rodrigues S P, Liu Y, Werner D H, and Cai W 2015 An active metamaterial platform for chiral responsive optoelectronics *Adv. Mater.* **27** p 4377
[20] Ronurpraful T, Jerop N, and Noginova N 2018 Enhancing sensitivity in plasmonic sensing *Materials Research Society Fall Meeting*
[21] Durach M, Rusina A, and Stockman M I 2009 Giant surface-plasmon-induced drag effect in metal nanowires *Phys. Rev. Lett.* **103** p 186801
[22] Durach M and Noginova N 2016 On the nature of the plasmon drag effect *Phys. Rev. B* **93** p 161406
[23] Durach M and Noginova N 2017 Spin angular momentum transfer and plasmogalvanic phenomena *Phys. Rev. B* **96** p 195411
[24] Strait J H, Holland G, Zhu W, Zhang C, Ilic B R, Agawal A, Pacifici D, and Lezec H J 2019 Revisting the photon-drag effect in metal films *Phys. Rev. Lett.* **123** p 053903
[25] Keene D, Ronurpraful T, and Noginova N 2019 Plasmon related electrical effects in strongly modulated metasurfaces *Proc. SPIE* **11080** 110801C
[26] Raether H, 1988 *Surface Plasmons on Smooth and Rough Surfaces and on Gratings* (New York, NY: Springer-Verlag)
[27] Li M, Cushing S K, and Wu N 2015 Plasmon-enhanced optical sensors: a review *Analyst* **140** pp 386-406
[28] Romanato F, Lee K H, Kang H K, Ruffato G, and Wong C C 2009 Sensitivity enhancement in grating coupled surface plasmon resonance by azimuthal control *Opt. Exp.* **17** p 2145
[29] Salabney A and Abdulhalim I 2012 Figure-of-merit enhancement of surface plasmon resonance sensors in the spectral interrogation *Opt. Lett.* **37** p 1175
[30] Eitan M, Iluz Z, Yifat Y, Boag A, Hanein Y, and Scheur J 2015 Degeneracy breaking of Wood's anomaly for enhanced refractive index sensing *ACS Photonics* **2** p 615
[31] Deng Y, Cao G, Yang H, Li G, Chen X, and Lu W 2017 Tunable and high-sensitivity sensing based on Fano resonance with coupled plasmonic cavities *Sci. Reports* **7** p 10639
[32] Johnson P B, and Christy R W 1972 Optical constants of the noble metals *Phys. Rev. B* **6** p 4370
[33] Rheims J, Köser J, and Wriedt T 1997 Refractive-index measurements in the near-IR using an Abbe refractometer *Meas. Sci. Technol.* **8** pp 601-605
[34] Hale G M, and Querry M R 1973 Optical constants of water in the 200-nm to 200-μm wavelength region *Appl. Opt.* **12** pp 555-563
[35] Sheldon M T, Van de Groep J, Brown A M, Polman A, and Atwater H A 2014 Plasmoelectric potentials in metal nanostructures *Science* **346** p 828

[36] Limonov M F, Rybin M V, Poddubny A N, and Kivshar Y S 2017 Fano resonances in photonics *Nat. Photon.* **11** p 543
[37] Kamenetskii E, Sadreev A, Miroshinichenko A, 2018 Fano Resonances in Optics and Microwaves (Heidelburg, Germany: Springer)
[38] Smith K C, Olafsson A, Hu X, Quillin S C, Idrobo J C, Collette R, Rack P D, Camden J P, and Masiello D J 2019 Direct Observation of Infrared Plasmonic Fano Antiresonances by a Nanoscale Electron Probe *Phys. Rev. Lett.* **123** p 177401
[39] Van de Groep J, Sheldon M T, Atwater H A, and Polman A 2016 Thermodynamic theory of the plasmoelectric effect *Sci. Reports* **6** p 23283
[40] Kurosawa H, Ohno S, and Nakayama K 2017 Theory of the optical-rectification effect in metallic thin films with periodic modulation *Phys. Rev. A* **95** p 033844